\documentclass{mem}
\usepackage{natbib}\usepackage{txfonts}\usepackage{balance}
\usepackage{graphicx}
\usepackage[a4paper,breaklinks,dvipdfm]{hyperref}
\idline{75}{282}

\begin{document}
\def\teff{$T\rm_{eff }$}
\def\kms{$\mathrm {km s}^{-1}$}

\title{ The Gamma Ray Bursts Hubble diagram}

   \subtitle{}

\author{
S. Capozziello\inst{1,2}, V.F. Cardone\inst{3},  M.G. Dainotti\inst{4}, \\ M. De Laurentis\inst{1,2}, L. Izzo\inst{5} M. \,Perillo\inst{2,6} 
          }

  \offprints{S. Capozziello}

\institute{ Dipartimento di Scienze
Fisiche, Universit\`{a} di Napoli "Federico II",
Complesso Universitario di Monte Sant'Angelo, Edificio N, via
Cinthia, 80126 - Napoli, Italy \and 
I.N.F.N. - Sez. di Napoli, Compl. Univ. Monte S. Angelo, Ed. G,
Via Cinthia, 80126 Napoli, Italy 
\and I.N.A.F. - Osservatorio Astronomico di Roma, via
Frascati 33, 00040\,-\,Monte Porzio Catone (Roma), Italy  
\and Dipartimento di Fisica and I.C.R.A., Universit\`{a} di Roma "La Sapienza", Piazzale Aldo Moro 5,  Roma, Italy
\and
Obserwatorium Astronomiczne, Uniwersytet Jagiello\'nski, ul. Orla 171, 31-501 Krak{\'o}w, Poland
\and
Dipartimento di Fisica ``E.R. Caianiello", Universit\`{a} di
Salerno, Via Ponte Don Melillo, 84081 Fisciano (Sa), Italy }

\authorrunning{S. Capozziello}

\titlerunning{ The Gamma Ray Bursts Hubble diagram}

\abstract{ Thanks to their enormous energy release, Gamma Rays
Bursts (GRBs) have recently attracted a lot of interest to probe
the Hubble diagram (HD) deep into the matter dominated era and
hence complement Type Ia Supernovae (SNeIa). We consider here
three different calibration methods based on the use of a fiducial
$\Lambda$CDM model, on cosmographic parameters and on the local
regression on SNeIa to calibrate the scaling relations proposed as
an equivalent to the Phillips law to standardize GRBs finding any
significant dependence. We then investigate the evolution of these
parameters with the redshift to obtain any statistical
improvement. Under this assumption, we then consider possible
systematics effects on the HDs introduced by the calibration
method, the averaging procedure and the homogeneity of the sample
arguing against any significant bias.}

\maketitle{}

\section{Introduction}

The observational evidences accumulated in the last
years, from the anisotropy and polarization spectra of the cosmic
microwave background radiation (CMBR),
the large scale structure traced by galaxy redshift surveys, 
the matter power spectrum with the imprints of the Baryonic Acoustic
Oscillations (BAO) and the Hubble diagram of
SNeIa, definitely
support the cosmological picture of a spatially flat universe with
a subcritical matter content $(\Omega_M \sim 0.3)$ and undergoing
a phase of accelerated expansion. From a theorical point of view,
the problem today is the presence of too many ideas, ranging from
the classical cosmological constant to scalar
fields  and higher order gravity theories all of them being more or less able
to fit the available data.
As often in science, adding further data and pushing the observed
Hubble diagram to higher redshift, calling into cause the so
energetic Gamma Ray Burst (GRBs), is the best strategy to put
order in this theoretical scenario. We believe that the existence
of many observationally motivated correlations, e.g.
\citep{Amati08}, offers the intriguing
possibility of turning GRBs into standardizeable candles just as
SNeIa.
Two main problems are actually still to be fully addressed. First,
all the  correlations have to be calibrated assuming a
fiducial cosmological model to estimate the redshift dependent
quantities. As a consequence, the so called circularity problem
comes out and we try to investigate if the different strategies
proposed to break it are viable solutions.
On the other hand, there is up to now no any definitive
understanding of the GRBs scaling relations so that one cannot
anticipate whether the calibration parameters are redshift
dependent. We address this question in a phenomenological way
adopting  different parameterizations.

\section{GRBs scaling relations}

To start, let us  consider first the general case of two observable
quantities $(x, y)$ related by a power\,-\,law relation which, in
a log\,-\,log plane, reads

\begin{equation}
\log{y} = a \log{x} + b  \ . \label{eq: loglog}
\end{equation}
Calibrating such a relation means determining the slope $a$, the
zeropoint $b$ and the scatter $\sigma_{int}$ of the points around
the best fit relation. Setting $y = \kappa d_L^2(z)$ with $\kappa$
a directly measurable redshift independent quantity and $d_L(z)$
the luminosity distance, one can then estimate the distance
modulus as\,:

\begin{eqnarray}
 \mu(z) & = & 25 + 5 \log{d_L(z)} \nonumber \\
~ & = & 25 + (5/2) (a\log{x} + b - \log{\kappa}) \label{eq: muval}
\end{eqnarray}

In order to perform such an estimate, one has to select a sample of
low redshift $(z \le 0.01)$ objects with known distance and fit
the scaling relation to infer the calibration parameters $(a, b,
\sigma_{int})$. Then, one has to assume that such calibration
parameters do not change with the redshift so that a measurement
of $(x, \kappa, z)$ and the use of the above scaling relation are
sufficient to infer the distance modulus. This approach, in principle, can be adopted for long and short GRBs \citep{felix}.

\subsection{2D empirical correlations}

We limit here our attention only to two dimensional (hereafter,
2D) correlations since they can be investigated relying on a
larger number of GRBs. These involve a wide range of GRBs
properties related to both the energy spectrum and the light curve
which are correlated with the isotropic luminosity $L$ or the
emitted collimation corrected energy $E_{\gamma}$. These last
quantities  depend on the luminosity distance $d_L(z)$ as
shown below:

\begin{equation}
L = 4 \pi d_L^2(z) P_{bolo} \ , \label{eq: lpbolo}
\end{equation}

\begin{equation}
E_{\gamma} = 4 \pi d_L^2(z) S_{bolo} F_{beam} (1 + z)^{-1}  \ ,
\label{eq: egamma}
\end{equation}
where $P_{bolo}$ and $S_{bolo}$ are the bolometric peak flux and
fluence, respectively, while $F_{beam} = 1 - \cos{(\theta_{jet})}$
is the beaming factor with $\theta_{jet}$ the rest frame time of
achromatic break in the afterglow light curve.
The combination of $x$ and $y$ gives rise to the different GRBs
correlations we will consider, namely the
$E_{\gamma}$\,-\,$E_{peak}$ \citep{G04}, the
$L$\,-\,$E_{peak}$ \citep{S03}, $L$\,-\,$\tau_{lag}$
\citep{N00}, $L$\,-\,$\tau_{RT}$ \citep{S07} and $L$\,-\,$V$
\citep{FRR00}.

\subsection{Bayesian fitting procedure}

Eq.(\ref{eq: loglog}) is a linear relation which can be fitted to
a given dataset $(x_i, y_i)$ in order to determine the two
calibration parameters $(a, b)$. The above linear relations will
be affected by an intrinsic scatter $\sigma_{int}$ which has to be
determined together with the calibration coefficients. To this
aim, in the following we will resort to a Bayesian motivated
technique \citep{Dago05}, which, however, does not tell us whether
this model fits well or not the data.

In order to sample the parameter space, we use a Markov Chain
Monte Carlo (MCMC) method running two parallel chains and using
the Gelman\,-\,Rubin (1992) test to check convergence \citep{CPC11}.

\subsection{GRBs luminosity distances}

A preliminary step in the analysis of  the 2D correlations
 is the determination of the luminosity $L$ or the
collimated energy $E_{\gamma}$ entering as $Y$ variable in the
$X$\,-\,$Y$ scaling laws with $(X, Y) = (\log{x}, \log{y})$. As
shown in Eqs.(\ref{eq: lpbolo})\,-\,(\ref{eq: egamma}), one has to
determine the GRBs luminosity distance over a redshift range
where the linear Hubble law does not hold anymore.
Different strategies have been developed to tackle this problem.
The simplest one is to assume a fiducial cosmological model and
determine its parameters by fitting, e.g., the SNeIa Hubble
diagram.  The $\Lambda$CDM  is usually adopted as fiducial model
thus setting\,:

\begin{equation}
E^2(z) = \Omega_M (1 + z)^3 + \Omega_{\Lambda} \label{eq: ezlcdm}
\end{equation}
with $\Omega_{\Lambda} = 1 - \Omega_M$ because of the spatial
flatness assumption. We determine the parameters $(\Omega_M, h)$,
using the Union2 SNeIa sample \citep{Union2} to get $(\mu_i^{obs},
\sigma_{\mu_i})$ for ${\cal{N}}_{SNeIa} = 557$ objects over the
redshift range $(0.015, 1.4)$ and set $(\omega_M^{obs},
\sigma_{\omega_M}) = (0.1356, 0.0034)$ for the matter physical
density $\omega_M = \Omega_M h^2$ and $(h, \sigma_h) = (0.742,
0.036)$ for the Hubble constant. The best fit values turn out to
be $(\Omega_M, h) = (0.261, 0.722)$.

Although the $\Lambda$CDM model fits remarkably well the data, it
is nevertheless worth stressing that a different cosmological
model would give different values for $d_L(z)$ thus impacting the
estimate of the calibration parameters $(a, \sigma_{int})$.
Looking for a model independent approaches, we first resort to
cosmography, i.e., we expand the scale factor
$a(t)$ to the fifth order and then consider the luminosity
distance as a function of the cosmographic parameters \citep{Izzo1,Izzo2}.

As a further step towards a fully model independent estimate of
the GRBs luminosity distances, one can use SNeIa as distance
indicator based on the naive observations that a GRBs at redshift
$z$ must have the same distance modulus of SNeIa having the same
redshift \citep{Izzo1}. Interpolating  the SNeIa Hubble diagram gives
 the value of $\mu(z)$ for a subset of the GRBs sample
with $z \le 1.4$ which can then be used to calibrate the 2D
correlations \citep{K08}. Assuming that this calibration
is redshift independent, one can  build up the Hubble diagram
at higher redshifts using the calibrated correlations for the
remaining GRBs in the sample. As in Cardone et al 2009,   we have used an approach based on the
local regression technique  which combines
much of the simplicity of linear least squares regression with the
flexibility of nonlinear regression.

\section{Calibration parameters}

While the $X$ quantities are directly observed for each GRB, the
determination of $Y$ (either the luminosity $L$ or the collimated
energy $E_{\gamma}$) needs for the object's luminosity distance.
The three methods described above allows us to get three different
values for $Y$ so that it is worth investigating whether this has
any significant impact on the calibration parameters $(a, b,
\sigma_{int})$ for the correlations of interest. We will refer
hereafter to the three samples with the $Y$ quantities estimated
using the luminosity distance from the fiducial $\Lambda$CDM
cosmological model, the cosmographic parameters and the local
regression method as the $F$, $C$ and $LR$ samples, respectively.
As a general result, we find that the fit is always quite good,
with reduced $\chi^2$ values close to $1$, in all the cases
independently of the 2D correlation considered and the distance
estimate method adopted.

The best fit coefficients and the median values clearly show that
the calibration based on the fiducial $\Lambda$CDM model leads to
steeper scaling laws for the most of  cases. On the contrary,
shallower slopes are obtained using the $C$ or $LR$ samples with
the $L$\,-\,$V$ relation as unique exception. Although the
differences in the slopes are not statistically meaningful because
of the large uncertainties, we find that the change in the slope
is not induced by the different luminosity distances adopted.

\section{Evolution with redshift}

It is not clear whether the calibration parameters $(a, b,
\sigma_{int})$ evolve with  redshift or not. To investigate
this issue, we consider two different possibilities for the
evolution with $z$. First, we consider the possibility that the
slope is constant, but the zeropoint is evolving. In particular,
we assume\,:

\begin{equation}
y = B (1 + z)^{\alpha} x^A \ \longrightarrow \ Y = \alpha \log{(1
+ z)} + a X + b \label{eq: plfit}
\end{equation}
with $(X, Y) = (\log{x}, \log{y})$ and $(a, b) = (A, \log{b})$.
Comparing  the previous constraints, we note that both the best
fit and median values of the slope parameter $a$ are significantly
shallower than in the no evolution case. However, the $68\%$
confidence ranges typically overlap quite well so that, from a
statistical point of view, such a result should not be overrated.
As such, we consider a most conservative option to assume that the
GRBs scaling relations explored here do not evolve with $z$.

As an alternative parametrization, we allow for an evolution of
the slope and not only the zeropoint of the 2D correlations. We
fit the data using\,:

\begin{equation}
Y = (a_0 + a_1 z) X + (b_0 + b_1 z) \ , \label{eq: linfit}
\end{equation}
i.e., we are Taylor expanding to the  first order the unknown
dependence of the slope and zeropoint on the redshift.
As a general result, we find that the best fit parameters and the
median values of the evolutionary coefficients $(\log{a_1},
\log{b_1})$ are typically quite small indicating that the
dependence of both the slope and the zeropoint on the redshift is
quite weak, if present at all.

\begin{figure}[]
\centering
\includegraphics[width=3cm]{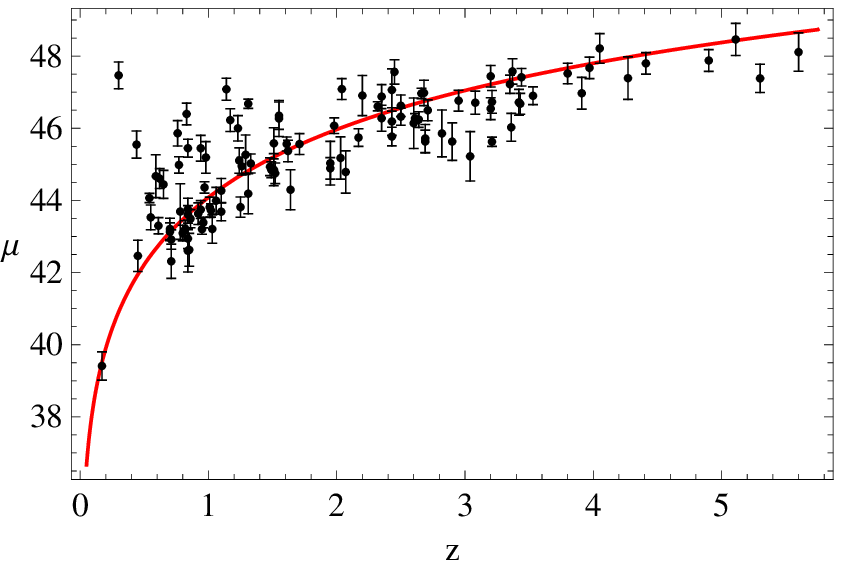}
\includegraphics[width=3cm]{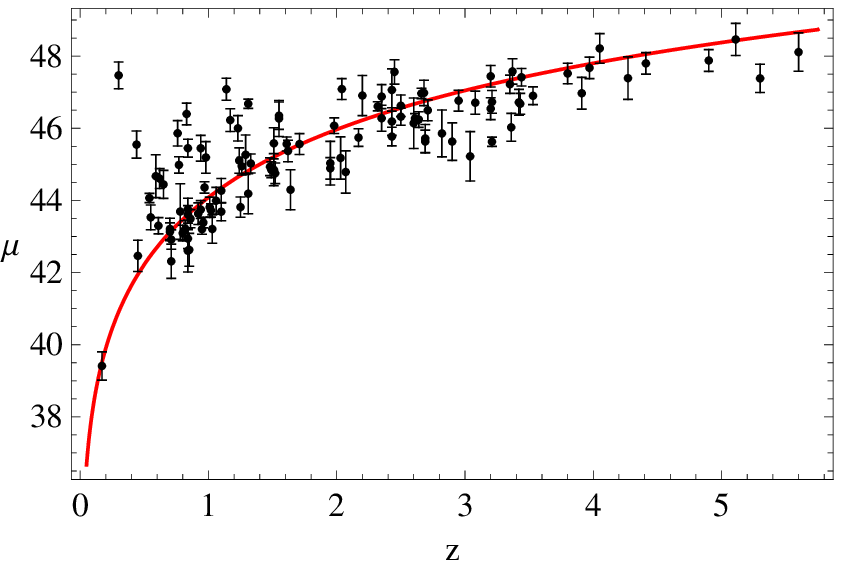}
\includegraphics[width=3cm]{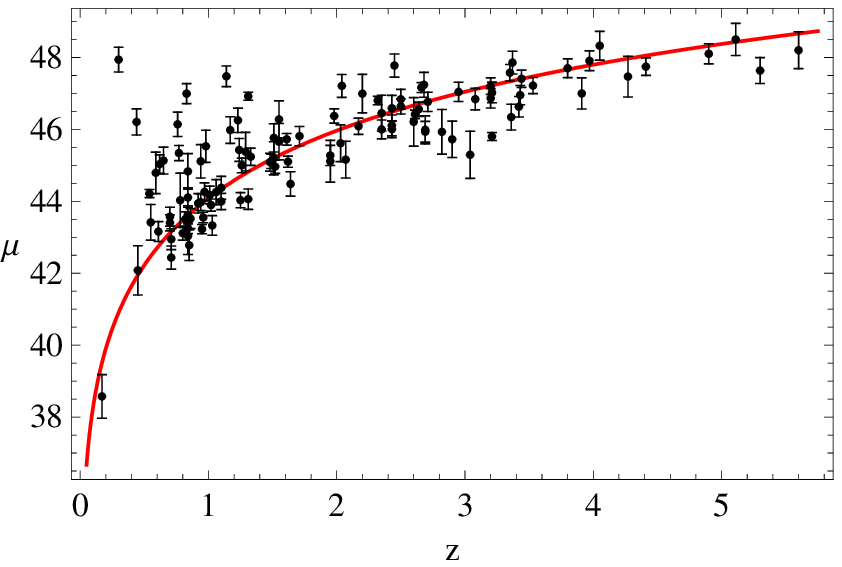} \caption{GRBs
Hubble diagrams (HDs) averaging over the six 2D correlations.
Three panels refer to the HDs derived using the calibration based
on the fiducial $\Lambda$CDM model (left), the cosmographic
parameters (right), and the local regression (down).} \label{fig:
grbhds}
\end{figure}

\section{GRBs Hubble diagram}

Once the calibration parameters for a given $Y$\,-\,$X$
correlation have been obtained, it is then possible to estimate
the distance modulus of a given GRB from the measured value of
$X$, as shown in Eqs.(\ref{eq: muval}), where $(a, b)$ are the
best fit coefficients for the given $Y$\,-\,$X$ correlation, while
$\kappa = 4 \pi P_{bolo}$, $\kappa = 4 \pi S_{bolo} F_{beam}/(1 +
z)$ and $\kappa = 4 \pi S_{bolo}/(1 + z)$ for $Y = L$, $Y =
E_{\gamma}$ and $E_{iso}$, respectively.
It is then possible to both reduce the uncertainties and
(partially) wash out the hidden systematic errors by averaging
over the different correlations available for a given GRB.

\subsection{Impact of the calibration method}

Fig.\,\ref{fig: grbhds} shows the GRBs Hubble diagrams (hereafter,
HDs) obtained averaging over the above 2D correlations and using the
three different calibration methods. The red solid line is the
expected $\mu(z)$ curve for the fiducial $\Lambda$CDM model.

As a general remark, we find that, notwithstanding the calibration
method adopted, the GRBs HDs reasonably follow the $\Lambda$CDM
curve although with a non-negligible scatter. Quite surprisingly,
the scatter is significantly larger in the range $0.4 \le z \le
1.4$ because of a set of GRBs with $\mu(z)$ lying systematically
above the $\Lambda$CDM prediction. One should argue for a failure
of the theoretical model, but there are actually a set of points
which are hard to reconcile with any reasonable dark energy model.

In order to compare the HDs from the three different calibration
methods, we consider the values of $\Delta \mu = \mu_{fid}(z) -
\mu(z)$ with $\mu_{fid}(z)$ the theoretically predicted distance
modulus for the fiducial $\Lambda$CDM model and then we conclude
that the HDs, obtained by using different calibration methods, are
consistent with each other within the uncertainties.

\subsection{Impact of the averaging procedure}

As yet stated above, averaging the $\mu$ values from different
correlations helps reducing the total uncertainties and partially
washes out the systematics connected to each single scaling
relations.

As a first check, we compare the $\Delta \mu$ values obtained
estimating $\mu$ using each single correlation. While the median
values of $\Delta \mu$ are roughly comparable, both $\langle
\Delta \mu \rangle$ and $(\Delta \mu)_{rms}$ are definitely larger
for the $L$\,-\,$E_{peak}$ and $L$\,-\,$V$ correlations. Pending
the question of which relation is physical, we can quantify the
impact of an incorrect assumption by evaluating again the distance
moduli excluding the $L$\,-\,$V$ and $L$\,-\,$E_{peak}$
correlations.

\subsection{Satellite dependence}

The  GRBs sample is made out by collecting the data available
in the literature so that the final catalog is not homogenous at
all. In order to investigate whether this could have any impact on
the HD, we consider again the deviations from the fiducial
$\Lambda$CDM model using only the 80 GRBs detected with the {\it
Swift} satellite. Somewhat surprisingly, we find larger $\Delta \mu$ values
independent of the calibration procedure adopted.

\section{Conclusions}

GRBs have recently attracted a lot of attention as promising
candidates to expand the Hubble diagram up to very high $z$.

As the Phillips law is the basic tool to standardize SNeIa, the
hunt for a similar relation to be used for GRBs has lead to
different empirically motivated 2D scaling relations. However, the
lack of a local GRBs sample leads to the so called {\it
circularity problem}. In an attempt to overcome this problem, we
have here considered the impact on the scaling relations and GRBs
HD of three different methods to estimate the luminosity distance,
concluding that they lead to consistent results.  The Hubble
diagrams averaging over the  correlations considered is not
affected by the choice of the calibration method.

Once the calibration procedure has been adopted, one has still to
check whether a redshift evolution of the GRBs scaling relations
is present or not.  We have therefore explored two different
parameterizations concluding that such an evolution is not
statistically motivated and it can be neglected.

Assuming that no evolution is present, we have finally checked
that the derived Hubble diagrams are not affected by systematics
related to the choice of the calibration method, the averaging
procedure or the homogeneity of the sample. As such, the GRBs HD
could be safely used as a tool to constrain cosmological
parameters.

\bibliographystyle{aa}

\begin{thebibliography}{}

\bibitem[{Amati et al. }{2008}]{Amati08}
Amati, L., Guidorzi, C., Frontera, F., Della Valle, M., Finelli,
F., Landi, R., Montanari, E. 2008, MNRAS, 391, 577

\bibitem[{Capozziello \& Izzo }{2008}]{Izzo1}
Capozziello, S., Izzo, L., 2008, A\&A 490,
31

\bibitem[{Capozziello et al. }{2011}]{felix}
Capozziello, S.,  De Laurentis, M., De Martino, I., Formisano, M.,  2011 Ast.Sp.Sc., 332, 31


\bibitem[{Cardone et al. }{2009}]{CCD09}
Cardone, V.F., Capozziello, S., Dainotti, M.G. 2009, MNRAS, 400,
775

\bibitem[{Cardone et al. }{2011}]{CPC11}
Cardone, V.F., Perillo, M, Capozziello, S., 2011, arXiv:1105.1122 [astro-ph.CO]  to appear in MNRAS

\bibitem[{D' Agostini }{2005}]{Dago05}
D' Agostini, G. 2005, arXiv\,:\,physics/051182


\bibitem[{Fenimore \& Ramirez\,-\,Ruiz }{2000}]{FRR00}
Fenimore, E.E., Ramirez\,-\,Ruiz, E. 2000, ApJ, 539, 712

\bibitem[{Ghirlanda et al. }{2004}]{G04}
Ghirlanda, G., Ghisellini, G., Lazzati, D. 2004, ApJ, 616, 331

\bibitem[{ Izzo et al }{2009}]{Izzo2}
 Izzo, L.,  Capozziello, S., Covone, G., Capaccioli, M. 2009, A\&A 508,
63

\bibitem[{Kodama et al. }{2008}]{K08}
Kodama, Y., Yonetoku, D., Murakami, T., Tanabe, S., Tsutsui, R.,
Nakamura, T. 2008, MNRAS, 391, L1

\bibitem[{Kowalski et al. }{2008}]{Union2}
Kowalski, M., Rubin, D., Aldering, G., Agostinho, R.J., Amadon, A.
et al. 2008, ApJ, 686, 749

\bibitem[{Norris et al. }{2000}]{N00}
Norris, J.P., Marani, G.F., Bonnell, J.T. 2000, ApJ, 534, 248


\bibitem[{Schaefer }{2003}]{S03}
Schaefer, B.E. 2003, ApJ, 583, L67

\bibitem[{Schaefer }{2007}]{S07}
Schaefer, B.E. 2007, ApJ, 660, 16


\end{thebibliography}

\end{document}